\documentclass[a4paper,twoside]{article}
\newcommand{\affil}[1]{$^{\rm #1}$}
\usepackage{amssymb}
\usepackage[authoryear]{natbib}
\bibpunct{(}{)}{;}{a}{}{,}
\usepackage{graphicx}
\date{} %Please leave the date blank
\title{\large\bf\flushleft Gravitational wave detection using pulsars: status of the Parkes Pulsar Timing Array project}
\author{\parbox{\textwidth}{\flushleft
\vspace{-0.5cm}
{\it G. B. Hobbs\affil{A}, M. Bailes,\affil{B}, N. D. R. Bhat\affil{B}, S. Burke-Spolaor\affil{A,B}, D. J. Champion\affil{A}, W. Coles\affil{C}, A. Hotan\affil{D}, F. Jenet\affil{E}, L. Kedziora-Chudczer\affil{A}, J. Khoo\affil{A}, K. J. Lee\affil{E,F}, A. Lommen\affil{G}, R. N. Manchester\affil{A}, J. Reynolds\affil{A}, J. Sarkissian\affil{A},  W. van Straten\affil{B}, S. To\affil{A,H}, J. P. W. Verbiest\affil{A,B}, D. Yardley\affil{A,H}, X. P. You\affil{I}}\\
\vspace{0.4cm}
{\small \affil{A}\,Australia Telescope National Facility, CSIRO, P.O. Box 76, Epping, NSW 1710 Australia.}\\
{\small \affil{B}\,Centre for Astrophysics and Supercomputing, Swinburne University of Technology, P.O. Box 218, Hawthorn VIC 3122, Australia} \\
{\small \affil{C}\,Electrical and Computer Engineering, University of California at San Diego, La Jolla, California, U.S.A}\\
{\small \affil{D}\,Department of Imaging and Applied Physics, Curtin University, Bentley, Western Australia, Australia}\\
{\small \affil{E}\,Center for Gravitational Wave Astronomy, University of Texas at
Brownsville, 80 Fort Brown, Brownsville, TX 78520}\\
{\small \affil{F}\,Department of Astronomy, Peking University, 5 Haidian Lu, Beijing, 100871, China}\\
{\small \affil{G}\,Franklin and Marshall College, 415 Harrisburg Pike, Lancaster, PA 17604, U.S.A.}\\
{\small \affil{H}\,Institute of Astronomy, School of Physics A29, The University of Sydney, NSW 2006, Australia}\\
{\small \affil{I}\,School of Physical Science and Technology, Southwest University, 2 Tiansheng Road, Chongqing 400715, China}
}}
\begin{document}
\twocolumn[
\begin{changemargin}{.8cm}{.5cm}
\begin{minipage}{.9\textwidth}
\vspace{-1cm}
\maketitle
%
%
%%%%%%%%%%%%%     ABSTRACT    %%%%%%%%%%%%%
%Abstract of no more than 200 words here.
\small{\bf Abstract:}

The first direct detection of gravitational waves may be made through observations of pulsars.  The principal aim of pulsar timing array projects being carried out worldwide is to detect ultra-low frequency gravitational waves ($f \sim 10^{-9}$--$10^{-8}$\,Hz). Such waves are expected to be caused by coalescing supermassive binary black holes in the cores of merged galaxies.  It is also possible that a detectable signal could have been produced in the inflationary era or by cosmic strings.  In this paper we review the current status of the Parkes Pulsar Timing Array project (the only such project in the Southern hemisphere) and compare the pulsar timing technique with other forms of gravitational-wave detection such as ground- and space-based interferometer systems.

%%%%%%%%%%%%%     KEYWORDS    %%%%%%%%%%%%%
\medskip{\bf Keywords: } gravitational waves---pulsars: general
% Please write all keywords in lower case. PASA uses the
% standard list of subject headings adopted by The Astrophysical Journal
% and available from http://www.journals.uchicago.edu/ApJ/keywords_text.html.
% Keywords are separated by em-dashes, i.e. ---

%%%%%%%%DO NOT EDIT%%%%%%%%%%%%
\medskip
\medskip
\end{minipage}
\end{changemargin}
]
\small
%%%%%%%%EDIT FROM HERE%%%%%%%%%%%%

\section{Introduction}
%Please see the PASA Style Guide for help with correct layout for your manuscript.
%Examples of tables and figures are given below.

Direct detection of gravitational waves (GWs) would be of huge importance to the physics and astrophysics communities. Detection would simultaneously verify our basic understanding of gravitational physics and usher in a new era of astronomy. Using models of galaxy merger rates and the properties of coalescing black holes, \cite{svc+08} and references therein predict the existence of an isotropic, stochastic, low-frequency GW background.  Various models of cosmic strings \citep{dv01,cbs96} and the inflationary era \citep{tur97,bb08} also predict the existence of a low-frequency GW background \citep{mag00}.  In the frequency range accessible by pulsar timing these backgrounds can be described by their characteristic strain spectrum, $h_c(f)$, which can be approximated as
\begin{equation}
h_c(f) = A_g\left(\frac{f}{f_{\rm 1 yr}} \right)^\alpha
\end{equation}
where $f_{\rm 1 yr} = 1/1$\,yr.  The dimensionless amplitude of the background, $A_g$, and the spectral exponent, $\alpha$, depend on the type of background \citep{jhv+06}. \cite{saz78} and \cite{det79} showed that these GW signals could be identified through observations of pulsars.

\begin{figure}
\includegraphics[width=6cm,angle=-90]{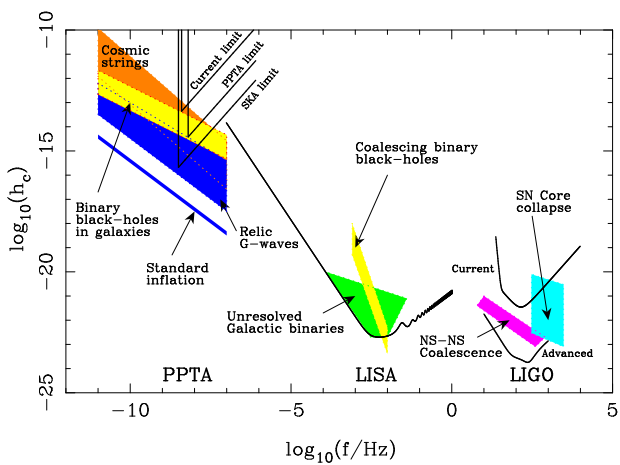}
\caption{Characteristic strain sensitivity for existing and proposed GW detectors as a function of GW frequency. Predicted signal levels from various astrophysical sources are shown. Figure from Hobbs (2008b).}\label{fg:sens}
\end{figure}

Standard pulsar timing techniques \citep[e.g.][]{ehm06,lk05} are used to search for the existence of GWs.  Timing software \citep{hem06} is used to compare the predictions of a pulsar timing model with measured pulse times of arrival (TOAs).  The timing model for each pulsar contains information about the pulsar's astrometric, spin and orbital parameters. Any deviations of the actual arrival times from the  model predictions - the pulsar timing residuals - represent the presence of unmodelled effects which will include calibration errors, spin-down irregularities and the timing signal caused by GWs.  The signal from GWs can be disentangled from other unmodelled effects by looking for correlations in the timing residuals of a large number of pulsars that are distributed over the entire sky. Pulsar timing experiments are sensitive to GW signals in the ultra-low frequency (f $\sim 10^{-9}$--$10^{-8}$\,Hz) band making them complementary to the space- and ground-based interferometers such as the Laser Interferometer Space Antenna (LISA) and the Laser Interferometer Gravitational Wave Observatory (LIGO), which are sensitive to higher frequency GWs (see Figure~\ref{fg:sens} and \S\ref{sec:band}).

\nocite{hob08b}

For a given pulsar and GW source, the induced timing residuals caused by a GW signal only depend on the GW characteristic strain at the pulsar and at the Earth.  The GW strains at the positions of well-separated pulsars will be uncorrelated, whereas the component at the Earth will lead to a correlated signal in the timing residuals of all pulsars \citep{hd83}.  It is this correlated signal that the pulsar timing experiments aim to detect.

The first high-precision pulsar timing array experiments were carried out using the Arecibo and GreenBank radio telescopes \citep[e.g.][]{fb90}.  Observations of PSRs~B1937+21 and B1855+09 were subsequently used to obtain a limit on the existence of a GW background \citep[e.g.][]{srtr90,ktr94,mzvl96,lom02}.  Unfortunately, the timing precision presented in these early papers was too poor to attempt to \emph{detect} the expected GW signals. New pulsar discoveries and better instrumentation has significantly improved the timing precision achievable.  Recently \cite{vbv+08} showed that PSR~J0437$-$4715 could be timed using the Parkes radio telescope with an rms timing residual of only 200\,ns over a 10\,yr period. Timing precisions are continuing to improve.

\cite{lb01} described how pulsar data sets could be used to constrain the existence of binary black holes in the centre of our Galaxy and in close-by galaxies. In 2003, a VLBI experiment detected motions in the radio galaxy 3C66B that were thought to be evidence of a binary supermassive black hole system \citep{simt03}.  \citet{jllw04} determined the expected PSR~B1855+09 timing residuals that would occur due to GW emission from the postulated system and compared them with archival Arecibo data. The pulsar timing observations allowed the existence of the binary system to be ruled out with 95\% confidence.

The Parkes Pulsar Timing Array (PPTA) project began in the year 2003.  The three main aims of this project are 1) to detect GW signals, 2) to establish a pulsar-based time scale for comparison with terrestrial time standards and 3) to improve the Solar system planetary ephemerides.  In this paper, we describe the work that has already been carried out towards our first aim of GW detection. This includes determining the number of pulsars and observing parameters required to detect a GW signal, obtaining the most stringent constraint on the existence of a GW background and producing high quality data sets for use in GW astronomy.  

\section{Current status of the Parkes Pulsar Timing Array project}

\begin{table*}[!t]
\caption{Observational parameters for current and future PPTA data sets. N$_p$ is the total number of observations, T$_s$ the data span, $\sigma$ the weighted rms timing residual over the entire data span. For the most recent data, $\sigma_{\rm med}^{\rm rec}$ gives the TOA uncertainty, S/N$_{\rm med}^{\rm rec}$ the median S/N and S/N$_{\rm max}$ the maximum S/N. S/N$_{\rm 100ns}$ is the predicted S/N required to produce a TOA uncertainty of 100\,ns and $\sigma_p$ is the predicted rms timing residual that may be possible with current instrumentation.}\label{tb:psrs}
\begin{center}\begin{tabular}{lrlllrrrl}\hline
PSR J  & N$_p$ & T$_s$ & $\sigma$ & $\sigma_{\rm med}^{\rm rec}$ & S/N$_{\rm med}^{\rm rec} $ & S/N$_{\rm max}^{\rm rec}$ & S/N$_{\rm 100ns}$ &$\sigma_{p}$ \\
             & & (yr) & ($\mu$s) & ($\mu$s) & & & & ($\mu$s) \\ \hline
J0437$-$4715 & 712 & 4.3 & 0.2 & 0.23 & 2271 & 2691 & 400  & 0.1$^\dagger$ \\
J0613$-$0200 & 432 & 5.5 & 1.1 & 0.75 & 123 & 226 & 917 & 0.5\\
J0711$-$6830 & 141 & 4.4 & 1.6 & 2.59 & 58 & 439 &  1699 & 1.0\\
J1022$+$1001 & 515 & 5.5 & 2.2 & 1.58 & 141 & 950 &  2301 & 0.5\\
J1024$-$0719  & 216 & 5.5 & 1.3 & 1.85 & 65 & 351 & 1219 & 1.0\\ \\
J1045$-$4509  & 155 & 5.2 & 3.0 & 2.08 & 158 & 240 & 3750 & 1.0\\
J1600$-$3053  & 567 & 5.5 & 1.0 & 0.44 & 157 & 297 & 693 & 0.3\\
J1603$-$7202  & 159 & 5.5 & 1.9 & 1.00 & 230 & 1462 & 2672 & 0.5\\
J1643$-$1224 & 186 & 5.4 & 1.7 & 0.63 & 324 & 500 & 2047 & 0.5\\
J1713$+$0747 & 303 & 5.5 & 0.5 & 0.20 & 270 & 702 & 584 & 0.1\\ \\
J1730$-$2304 & 158 & 4.6 & 1.9 & 1.08 & 212 & 459 & 2290 & 1.0\\
J1732$-$5049 & 188 & 5.5 & 3.5 & 2.12 & 76 & 118 & 1688 & 1.0\\
J1744$-$1134 & 413 & 5.5 & 0.8 & 0.67 & 79 & 252 & 693 & 0.3\\
J1824$-$2452 & 169 & 3.1 & 1.7 & 1.57 & 18 & 43 & 323 & 1.0\\
J1857$+$0943 & 205 & 4.4 & 1.4 & 0.91 & 115 & 415 & 1087 & 0.5\\ \\
J1909$-$3744 & 1398 & 5.5 & 0.6 & 0.16 & 127 & 620 & 204 & 0.1\\
J1939$+$2134 & 45 & 4.3 & 0.3 & 0.19 & 69 & 222 & 77 & 0.1\\
J2124$-$3358 & 125 & 3.8 & 2.4 & 2.81 & 61 & 205 & 2000 & 1.0\\
J2129$-$5721 & 189 & 5.5 & 1.2 & 2.75 & 47 & 252 & 1234 & 1.0\\
J2145$-$0750 & 205 & 4.3 & 1.1 & 0.81 & 275 & 1665 & 2346 & 0.3\\
\hline\end{tabular}\end{center}
\footnotesize{$^\dagger$ The timing residuals for PSR~J0437-4715 are currently dominated by systematic effects. Understanding and removing of these effects may allow us to reach a TOA timing precision of $\sim 20$\,ns; see text.}
\end{table*}

The PPTA project has been described in numerous papers \citep[for an historical overview see][]{hob05,man06,hob08,man08}. In brief, the 20 millisecond pulsars listed in Table~\ref{tb:psrs} are observed with an $\sim2$-week cadence using the 64-m Parkes radio telescope.  The pulsars were chosen because they 1) only exhibit small-scale timing irregularities, 2) have been well studied by earlier projects and 3) have a range of separations on the sky.  Observations are carried out with a 20cm receiver (providing 256\,MHz of bandwidth) and with a 10cm/50cm dual-band system (giving 64MHz of bandwidth at 50cm and 1GHz of bandwidth at 10cm).  Multiple backend systems are used including digital filterbanks and the Caltech-Parkes-Swinburne Recorder 2 \citep[CPSR2;][]{hbo06} which provides two 64\,MHz bands of coherently dedispersed data.  New backends that are currently being commissioned should further improve our timing precision.  All our observations are processed using the \textsc{psrchive} software \citep{hvm04} and pulse TOAs are analysed using \textsc{tempo2} \citep{hem06,ehm06}.

\subsection{The initial data sets}

The choice of 20 pulsars was based on work reported by \cite{jhlm05} who determined the minimum number of pulsars required to detect the expected GW background. This work highlighted the requirement for extremely high precision timing observations - in order to detect the GW background within five years all 20 pulsars need to have rms timing residuals of $\sim 100$\,ns.  \cite{yhc+07}  showed that the interstellar medium and Solar wind can produce much larger signals in the timing residuals. For instance, at an observing frequency close to 1.4\,GHz, the Solar wind can produce delays of $\sim 100$\,ns for lines-of-sight to the pulsar that pass within 60$^\circ$ of the Sun and much larger delays for closer lines-of-sight \citep{yhc+07a}. Fortunately dispersive effects can be removed by comparing pulse arrival times at different observing frequencies. An optimal subtraction technique was presented by \cite{yhc+07}.

Combining the initial PPTA data sets (that spanned $\sim$2\,yr) with archival Arecibo observations allowed \cite{jhv+06}  to obtain the most stringent limit to date on the amplitude of a GW background of $A_g \leq 1.1\times10^{-14}$. This corresponds to $\Omega_g[1/(8{\rm yr})]h^2 \leq 1.9\times10^{-8}$ for the fractional energy density per unit logarithmic frequency interval for an astrophysical background (the Hubble parameter $H_0 = 100h$\,km\,s$^{-1}$\,Mpc$^{-1}$). This result constrained the merger rate of supermassive binary black-hole systems, ruled out some relationships between the black-hole mass and the galactic halo mass, constrained the rate of expansion in the inflationary era and provided an upper bound on the dimensionless tension of a cosmic-string background.  Unfortunately, the technique of \cite{jhv+06} is only valid for data sets that have a ``white spectrum'' (defined as being a spectrum that is independent of frequency).  The initial data sets, and previously published work \citep[e.g.][]{ktr94,cb04,sns+05,vbv+08}, clearly show that many millisecond pulsar timing residuals are affected by unexplained timing irregularities which may be related to the timing noise phenomena seen in younger pulsars \citep{hlk06}, unmodelled instrumental effects, poor removal of interstellar medium effects \citep[e.g][]{sti06} or intrinsic pulse shape variability.  This limitation restricted our initial analysis to only seven of the 20 PPTA pulsars.  We have recently developed a new technique that makes no assumption on the spectrum of the timing residuals.  This work will be presented in a forthcoming paper.

\section{Towards detecting GW signals}

Timing residuals for our current PPTA data sets are summarised in Table~\ref{tb:psrs}.  To form these data sets we have selected the backend systems providing most precise TOAs.  The earliest data are usually from the CPSR2 coherent dedispersion system and the most recent data from the digital filterbank systems.  Columns 2, 3 and 4 provide the total number of observations, the time span and the weighted rms timing residual over our entire data span respectively (obtained after fitting only for the pulsar's astrometric and orbital parameters along with the pulse frequency and its first derivative). These timing residuals have not been corrected for the effects of dispersion measure variations caused by the interstellar medium.  Note also that multiple, independent observations at different frequencies have been made during the same observing session.  A weighted average (in 2-weekly intervals) of the residuals for PSR~J0613$-$0200 gives an rms of 842\,ns with 102 data points (cf. 1.1$\mu$s listed in the table).  As our observing systems have been improved over this period, the median arrival-time uncertainties for our most recent observations are listed in Column 5 of Table~\ref{tb:psrs}. 

\begin{figure}
\includegraphics[angle=-90,width=8cm]{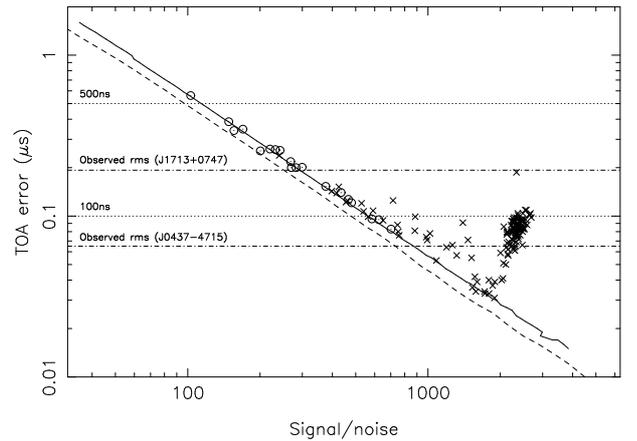}
\caption{Arrival time precision versus signal to noise ratio for PSRs~J1713+0747 and J0437$-$4715. For PSR~J1713$+$0747 the measured values are plotted as circles and a theoretical estimate is drawn as the solid line. For PSR~J0437$-$4715 we use `cross' symbols and a dashed line respectively.   The observed weighted rms timing residuals are indicated by dashed lines. The dotted lines represent timing precisions of 100\,ns and 500\,ns.}\label{fg:1713}
\end{figure}

Even though many of these data sets have lower rms values than have been achieved in the past, we are still far from the goal of $\sim 100$\,ns for the majority of our sample.   For a given backend system it is possible to use a noise-free approximation of the pulse profile to determine the best timing precision achievable for a given pulsar.  This can be carried out through simulations or by using analytical expressions \citep{van06,dr83}.  A typical example is shown in Figure~\ref{fg:1713} where the actual arrival time uncertainties for PSR~J1713$+$0747 are compared with the theoretical prediction. For this pulsar it is clear that we should be able to achieve typical arrival time uncertainties $\sim 100$\,ns.  PSR~J0437$-$4715 provides our best quality data sets.  However, as shown in Figure~\ref{fg:1713} we are far from the theoretical timing precision achievable of $\sim 20$\,ns for this pulsar because of un-explained systematic effects. In Table~\ref{tb:psrs} we list the signal-to-noise (S/N) ratio of an individual observation that would be required in order to measure an arrival time with a precision of 100\,ns (S/N$^{\rm rec}_{\rm 100ns}$). The table also contains the median (S/N$^{\rm rec}_{\rm med}$) and maximum (S/N$^{\rm rec}_{\rm max}$) S/N ratios achieved to date using our most recent observing system.  Further improvements should be possible by improved calibration, using a new coherent dedispersion system currently being commissioned, longer observation lengths and by obtaining pulse times-of-arrival using the information available from the Stokes parameters \citep{van06} instead of considering only the total intensity profile.

\begin{figure}
\includegraphics[angle=-90,width=8cm]{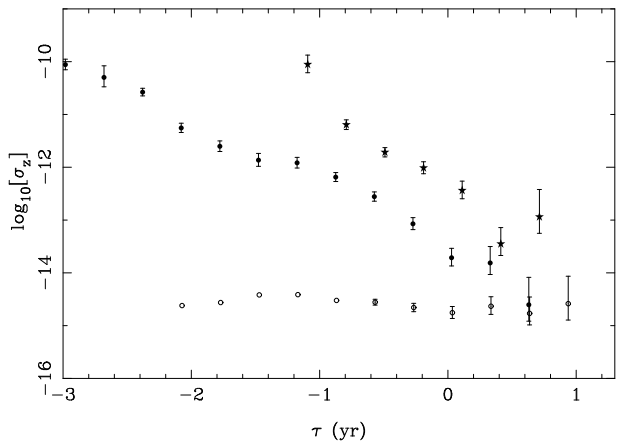}
\caption{The $\sigma_z(\tau)$ stability parameter for PSRs J0437$-$4715 (solid circle) and J1045$-$4509 (star symbols).  For comparison the difference between two realisations of the terrestrial time scale is also shown (open circles). Error bars are only plotted if they are larger than the size of the data point.}\label{fg:sigmaz}
\end{figure}

In order to achieve rms timing residuals of $\sim$100\,ns over five years of observations it is necessary that the pulsars are intrinsically stable over this time scale.  A generalisation of the Allan variance has been developed for analysing pulsar timing residuals \citep{mte97}. This measure of stability as a function of time-span,  $\sigma_z(\tau)$, has been plotted in Figure~\ref{fg:sigmaz} for our best-timed pulsar, PSR~J0437$-$4715, and for one of our poorer pulsars, PSR~J1045$-$4509. The figure also includes measurements of $\sigma_z$ for the difference between the realisations of terrestrial time published by the National Institute of Standards and Technology (NIST) and the Physikalisch-Technische Bundesanstalt (PTB).  These initial results demonstrate that our best pulsars are stable enough to search for irregularities in the terrestrial time scales and could be used to form a pulsar-based time scale.     

\subsection{The GW background}

\begin{figure}
\includegraphics[angle=-90,width=8cm]{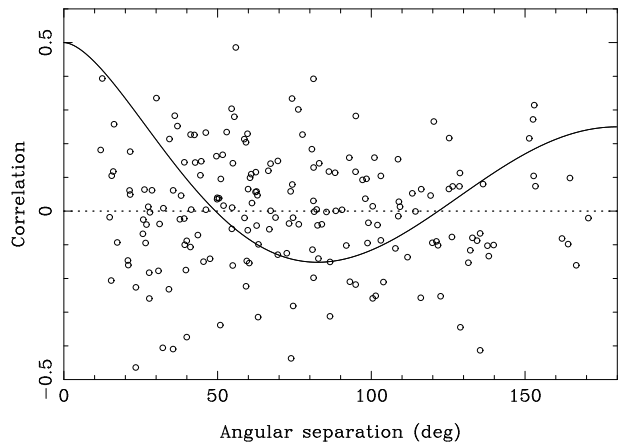}
\caption{Correlation between all pairs of pulsars plotted versus their angular separation. The solid line is the predicted correlation curve due to a GW background.}\label{fg:corr}
\end{figure}

A GW background will induce timing residuals that are correlated between different pulsars. The general relativistic prediction for this correlation is shown as the solid line in Figure~\ref{fg:corr} \citep{hd83}.  Determining the correlation coefficients for our data sets is not trivial due to the uneven sampling, differing data lengths and the presence of low-frequency noise. However, an initial analysis in which we averaged the measured timing residuals into $2$-weekly sessions before using standard correlation techniques, produced the correlation coefficients shown in Figure~\ref{fg:corr}. Clearly we have not yet made a significant detection.  

A reference timing array project was defined by \cite{jhlm05} containing 20 pulsars, timed with an rms timing residual of 100\,ns with weekly sampling over 5 years.  Such observations would give a detection significance of $\sim 3$ for a background with $A_g = 2 \times 10^{-15}$. With our current rms timing residuals and fortnightly sampling we would require data spanning more than 16\,yr in order to achieve this sensitivity.  For rough estimates of the timing precision that we are likely to achieve with improved instrumentation (listed in column 9 in Table~\ref{tb:psrs} and obtained by postulating possible future S/N ratios and their corresponding TOA errors), we require data spans of $\sim$10\,yr to reach this sensitivity.  

In order to reduce the time required to create such data sets, we are developing collaborations with the North American\footnote{http://nanograv.org}, European \citep{jsk+08,skl+06} and Chinese \citep{wzzz02} timing array projects.  It is also expected that on-going pulsar surveys will discover new, stable pulsars that can be used in future timing array projects.  In the longer term we expect that the Square Kilometre Array (SKA) telescope \citep{cr04,ckl+04,kra04b} will provide high quality data sets for many hundreds of pulsars.  Such observations will allow the properties of the GW background to be studied in detail.  For instance, predictions of various theories of gravity for the polarisation properties of GWs could be tested (Lee et al., in press).

\subsection{Single GW sources}

Single GW sources can be divided into individual supermassive black-hole binary systems and burst GW sources.  Order-of-magnitude calculations demonstrate that a system with a chirp mass of $10^9 {\rm M}_\odot$ and an orbital period of 10\,yr at a distance of 20\,Mpc will induce a sinusoidal signal in the timing residuals with amplitude $\sim 100$\,ns at a frequency of twice the orbital frequency of the binary black holes.  More detailed calculations, which take various cosmological effects into account when determining our sensitivity to single GW sources, will be published shortly.  This new work demonstrates that our data sets can be used to place constraints on the merger rate of supermassive binary black-hole systems as a function of black-hole mass and redshift. 

Sources of burst GW emission that may be detectable with current and proposed pulsar timing experiments include 1) the formation of supermassive black holes \citep{tb76}, 2) highly eccentric supermassive black hole binaries \citep{en07}, 3) close encounters of massive objects \citep{kgm06} and 4) cosmic string cusps \citep{dv01}.  Our ability to detect and localise the position of burst sources  depends on our having accurate data sets for a large number of pulsars. By sharing data with timing array projects in the Northern hemisphere we expect to improve both the angular resolution of our detector and our sensitivity to burst sources.

\section{The pulsar timing GW band}\label{sec:band}

Any GW signal with a period longer than the time-span of the data is largely absorbed by the fitting process. Pulsar timing is therefore only sensitive to GW signals with periods less than the time-span of the data ($f  \lesssim 10^{-8}$\,Hz). The planned space-based GW detector, LISA, is sensitive to higher frequency GWs ($f \sim 10^{-3}$\,Hz) and the ground based interferometers  to even higher frequencies ($f \sim 10^2$\,Hz).  In general, the expected GW sources are different for each type of detector.  However, some models of cosmic strings and the inflationary era produce GW backgrounds which could, potentially, be detectable at all GW observing frequencies \citep{mag00}.  

As noted by \cite{svc+08}, detection of a background from coalescing binary black hole systems by both the pulsar timing array experiments and LISA would provide strong constraints on models of the assembly of massive binary black hole systems.  Recently, \cite{pch+08} discussed other areas where pulsar timing and LISA may provide complementary observations.  For instance, they consider the possibilities of studying the inspiral stage of a merger using pulsar observations and the ring-down stage using LISA.

\section{Conclusions}

Current pulsar timing array experiments have the potential to detect low-frequency GW signals within a decade.  Future experiments with the Square Kilometre Array telescope should allow detailed observations of both a stochastic GW background and individual GW sources.  The pulsar timing technique is complementary to other projects that are attempting to detect much higher frequency GWs.

\section*{Acknowledgments} %If needed

This work is undertaken as part of the Parkes Pulsar Timing Array project. The Parkes radio telescope is part of the Australia Telescope, which is funded by the Commonwealth of Australia for operation as a National Facility managed by the Commonwealth Scientific and Industrial Research Organisation (CSIRO). This research was funded in part by the National Science Foundation (grant \#0545837) and RNM's Australian Research Council Federation Fellowship (project \#FF0348478). GH is the recipient of an Australian Research Council QEII Fellowship (\#DP0878388). XPY is supported by the Natural Science Foundation Project CQ CSTC 2008BB0265.

%\begin{thebibliography}{}
% References are listed as in the following example, for more examples, please
% see the PASA Style Guide

%\begin{thebibliography}{}
% References are listed as in the following example, for more examples, please
% see the PASA Style Guide
%\include{ref}
%\bibitem[Smith, Jones, \& Brown(Year)Smith et al.]{example}Smith, A.~B., Jones,
%C.~D., Brown, E.~F. Year, Journal, Volume, Page
%\end{thebibliography}

%\end{thebibliography}

\end{document}